\documentclass[english,aps,preprint]{revtex4}
\usepackage[T1]{fontenc}
\usepackage[utf8]{luainputenc}
\usepackage{amsmath}
\usepackage{graphicx}
\usepackage{esint}

\makeatletter
\@ifundefined{textcolor}{}
{%
 \definecolor{BLACK}{gray}{0}
 \definecolor{WHITE}{gray}{1}
 \definecolor{RED}{rgb}{1,0,0}
 \definecolor{GREEN}{rgb}{0,1,0}
 \definecolor{BLUE}{rgb}{0,0,1}
 \definecolor{CYAN}{cmyk}{1,0,0,0}
 \definecolor{MAGENTA}{cmyk}{0,1,0,0}
 \definecolor{YELLOW}{cmyk}{0,0,1,0}
}

\makeatother

\usepackage{babel}
\begin{document}

\title{Electron acceleration by coherent laser pulse echelons in periodic
plasma structures}

\author{A. Pukhov$^{1,2}$, I. Kostyukov$^{2,3}$, T. Tückmantel$^{1}$,
Ph. Luu-Thanh$^{1}$}

\address{$^{1}$University of Dusseldorf, 40025 Germany}

\address{$^{2}$University of Nizhny Novgorod, Nizhny Novgorod 603950, Russia }

\address{$^{3}$Institute of Applied Physics RAS, Nizhny Novgorod 603950,
Russia }
\begin{abstract}
We consider a possibilty to use an echelon of mutually coherent laser
pulses generated by the emerging CAN (Coherent Amplification Network)
technology for direct particle acceleration in periodic plasma structures.
The plasma structure survives a single shot only. However, due to
it's simplicity and projected very low production costs, the structure
can be replaced for every laser shot at a kiloherz repetition rate.
We discuss resonant and free streaming configurations. The resonant
plasma structures can trap energy of longer laser pulses but are limited
to moderate laser intensities of about $10^{14}\,{\rm W/cm^{2}}$
and are very sensitive to the structure quality. The free streaming
configurations can survive laser intensities above $10^{18}\,{\rm W/cm^{2}}$
for several tens of femtoseconds so that sustained accelerating rates
well above ${\rm TeV/m}$ are feasible. In our full electromagnetic
relativistic particle-in-cell (PIC) simulations we show a test electron
bunch gaining up to $120\,{\rm GeV}$ over a distance of $5.3\,{\rm cm}$
only.
\end{abstract}
\maketitle

\section{introduction}

More than twenty years ago the invention of chirped-pulse-amplification
(CPA) technique triggered a revolution in laser technology and lead
to ultrashort laser pulses with huge peak powers \cite{Mourou1985}.
The novel laser sources allowed for a steady advance in a number of
important scientific and industrial applications: high-field physics,
``attosecond'' physics and high-order harmonic generation, secondary
compact sources of particles and radiation, inertial confinement fusion,
probing and imaging of ultrafast processes for biology, medicine and
material sciences, etc. The impressive progress has been demonstrated
in plasma-based acceleration methods with ultrashort high-power laser
pulses used as a driver generating accelerating plasma structures.
Quasimonoenergetic electron bunches with the energy up to $1$ GeV
have been generated in the pioneering experiments \cite{Faure2004,Mangles2004,Geddes2004,Leemans2006}.
Now the $1$ GeV energy barrier has been broken and electron bunches
with a few percent energy spread and unprecedented sub-milliradian
divergences have been recorded at $2$ GeV energy \cite{Downer2013}.
In these experiments, the electrons have been accelerated in underdense
plasmas in the so called bubble regime of laser wake fields \cite{Bubble}.
The traditional CPA lasers use TiSa crystals and deliver an average
power of a few Watts only and a low efficiency. This is a significant
drawback when we think about high energy physics applications where
high luminosities are usually required.

Another laser technique based on fiber amplifiers pushes the laser
average power up to kilowatt level and the laser wall-plug efficiency
to over $30\%$. These values are several orders of magnitude better
than the traditional TiSa lasers could possibly provide \cite{Jeong2004,Liem2004}.
The high peak power required by many applications can be achieved
when the fiber laser systems are combined with CPA pulse compression
and massive parallelism of multiple fiber amplifiers \cite{Ediam2011,Prawiharjo2008}. 

The real revolution however is promised by the CAN (Coherent Amplification
Network) technology \cite{Mourou2013}. Here, the high average power
and efficiency of the fiber laser systems is accomplished by the mutual
coherence of individual laser pulses \cite{Bellanger2010} in the
massively parallel fiber array. For the first time we are going to
have a huge number of identical - and thus mutually coherent - laser
pulses that can be flexibly arranged. It is expected that the laser-driven
particle acceleration schemes will strongly benefit from the CAN project
\cite{Mourou2013}. The coherency allows to control not only the space-time
distribution of the laser intensity in the acceleration region but
also the phase distribution. This promises to make the acceleration
much more robust and efficient. 

At the same time, the continuous progress in micro- and nanotechnology
stimulates exploration of dielectric-based laser-driven acceleration
structures \cite{Rosenzweig1995}. Such structures transform the transverse
laser field into accelerating and focusing fields suitable for efficient
particle acceleration. Various accelerating schemes have been suggested:
traveling wave structures \cite{Eddie2001}, semi-open structures\cite{Huang1996},
resonant-closed structures \cite{Yoder2005}, photon crystal structures
\cite{Tien1999}, phase-modulation-mask structures \cite{Plettner2006}
etc. However, all these structures were designed for a single laser
pulse driver as the traditional laser technology provides coherence
for a single laser pulse only. 

Here we show that the CAN technology may become a real game changer
in the structured acceleration. We propose several schemes that combine
the advantage of multiple coherent laser pulses with advantages of
resonant-closed and phase-modulation-mask structures. We concentrate
here on single-shot regime when the solid elements of the structure
are converted to plasma by the field ionization and evaporate after
the interaction. Yet we assume that the structure keeps its integrity
during the short laser pulse duration. Thus, we generate a regular
overdense plasma structure. In contrast to the previous dielectric-based
schemes \cite{Yoder2005,Plettner2006} we analyze robust schemes which
do not rely on the laser phase shift in solid elements of the structure
because this would be difficult to control in plasma. 

The regular plasma structures allow for the first time to use the
huge laser field itself and not a small perturbation generated by
the laser in plasma. The sustained accelerating field can thus reach
unprecedented values up to several TV/m.

\section{ACCELERATING STRUCTURE BASED ON HALF-PERIOD FABRY-PEROT PLASMA RESONATORS}

\begin{figure}
\includegraphics[width=0.95\columnwidth]{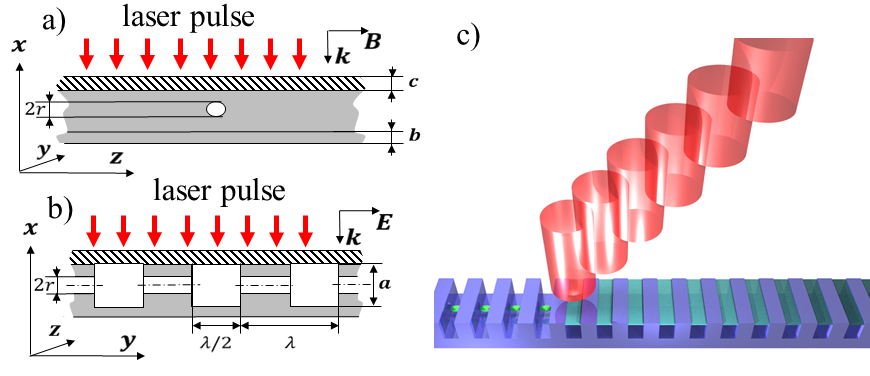} \caption{Schematic plot of the resonator accelerating structure: a) cut across
the witness beam axis; b) cut along the witness electron beam axis;
c) three dimensional view of the periodic resonator structure. Individual
laser pulses in the echelon are phased to each other and arrive at
the structure with a delay corresponding to the particle travelling
time. The dashed area in the two dimensional cuts is the semi-transparent
thin foil; the grey area is the overdense plasma elements; the white
areas are vacuum microcavities and the microchannel for the particle
beam transport. }

\label{fig1} 
\end{figure}

First we propose an accelerating structure based on a periodic set
of Fabry-Perot resonators. The structure consists of two-dimensional
resonator boxes on top of a solid substrate. The both sides of the
cavities are equal to half a laser wavelength. The cross section of
the structure geometry is shown in Fig.\ref{fig1}. The cavities are
covered by a thin foil (the dashed area in Fig. 1) that is partially
transparent for laser radiation while the other cavity walls reflect
radiation. The microcavities (the white quadratic areas in Fig. 1b)
may provide field enhancement up to the factor 

\begin{equation}
\frac{E_{{\rm c}}}{E_{0}}\approx\frac{1}{1-R}\label{eq:FP ampl}
\end{equation}
where $E_{c}$ is the field in the cavity, $E_{0}$is the laser field,
$R$ is reflectivity of the covering foil. It has been demonstrated
previously \cite{Shen2002} that intensity in the plasma cavity may
reach a $100$-fold that of the pump lasers. 

The field amplification \eqref{eq:FP ampl} is important for laser
pulses that are several periods long. The pulse energy gets trapped
in the resonator and can be harvested by the witness beam load. The
pulse duration $\tau_{p}$ and the optimal reflectivity of the semitransparent
foil are connected as 

\begin{equation}
\tau_{p}\approx\frac{\pi\lambda}{2c\left(1-R\right)}\label{eq:FP lifetime}
\end{equation}
where $\lambda$ is the laser wavelength. 

To achieve a phase synchronism between accelerated particles with
the velocity $\upsilon$ and the field, the microcavities must be
periodically intermitted along the beam axis with the period 

\begin{equation}
\Lambda=\lambda\frac{\upsilon}{c}\label{eq:period}
\end{equation}
Thus, we introduce overdense plasma hills of $\lambda/2$ width and
$\lambda$ period to screen the laser field in its decelerating phase
(the grey parts in Fig 1). The screening plasma hills should have
a hole along the structure axis for particle beam transportation.
The structure is irradiated from the $+x$ direction by a laser polarized
in the $y$ direction. The relativistic electron moves along the $y$
axis through the resonator cavity and through the screening elment
within the hole of radius $r$. An infinite extent in the $z$ dimension
is assumed. 

The laser pulse must be short enough and the laser intensity must
be moderate so that the structure keeps its integrity during the interaction.
Our PIC simulations using the code VLPL (Virtual Laser Plasma Lab)
\cite{VLPL} show that the plasma resonator structures may survive
intensities up to about $10^{14}{\rm W/cm^{2}}$, corresponding to
accelerating fields above $10\,{\rm GV/m}$.

\noindent 

The principle of the proposed accelerating structure is rather simple:
enhancement of the accelerating fields and damping of the decelerating
fields. The relativistic electron with a proper phase harvests the
accelerating field in the cavity during the half laser period. When
the laser field changes its sign and becomes decelerating, electrons
are screened from the field under the overdense plasma hill. Relativistic
electrons passing through the structure at a proper phase gain energy. 

\begin{figure}
\includegraphics[width=0.95\columnwidth]{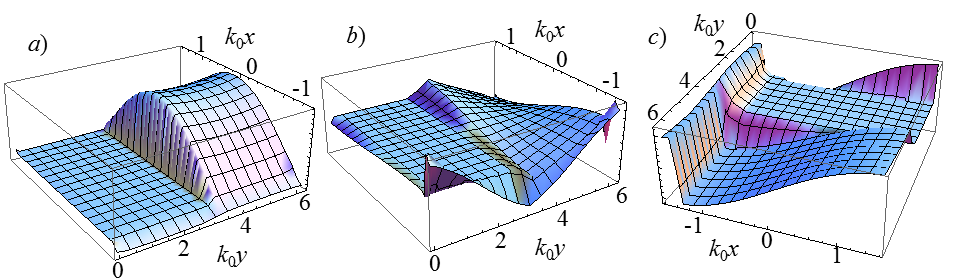}

\caption{Snapshot of the fields (a) $E_{y}$ with the normalized magnitude
$E_{y}=0.03$; (b) $E_{x}$ with the normalized magnitude $E_{x}=0.005$
and (c) $B_{z}$ with the normalized magnitude $B_{z}=0.005$. The
fields are shown inside the cavity and in the damping part of the
structure at the phase when the accelerating field $E_{y}$ reaches
its maximum.}

\label{fig2-1} 
\end{figure}

In our PIC simulations, we took a laser pulse with a Gaussian temporal
profile $a(t)=a_{0}\exp\left(-t^{2}/2\tau_{p}^{2}\right)$ with the
relativistic amplitude $a_{0}=eE_{L}/(mc\omega_{L})=0.01$ and duration
$\tau_{p}=10\lambda/c$. Here $\lambda$=800nm is the laser wavelength,
$E_{L}$ is the laser field amplitude, $\omega_{L}$ is the laser
frequency, $m$, $e$ and $c$ are the electron mass, the electron
charge and the speed of light, respectively. The plasma electron density
was assumed to be $n_{e}=4\times10^{23}$~cm$^{-3}$, the thickness
of the semitransparent foil was $d=0.02\lambda$. The longitudinal
electric field distribution calculated in the cavity and in the damping
part is shown in Fig.\ref{fig2-1}a. It is seen from Fig.\ref{fig2-1}a
that the electric field is amplified by a factor about $3$ in the
cavity due to the Fabry-Perot effect while it is supressed in the
damping part nearly completely.

\begin{figure}
\includegraphics[width=0.9\columnwidth]{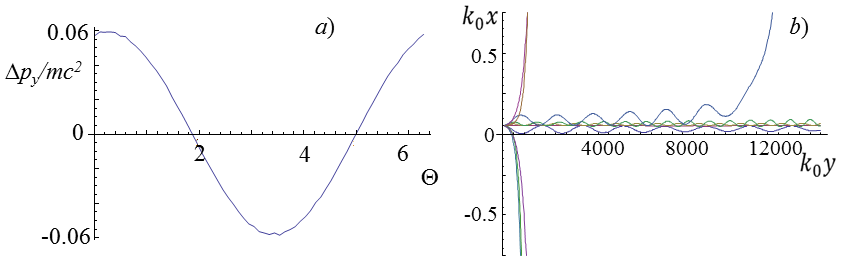}

\caption{a) Gain in the logitudinal momentum $\Delta p_{y}$ of an electron
passing through the one-period structure as function of the initial
field phase $\Theta$; b) trajectories of electrons injected at the
center of the cavity starting with $\gamma=200$, and evenly distributed
phases in the field phase $\Theta$ between $0$ and $2\pi$.}

\label{fig3} 
\end{figure}

We study the dynamics of relativistic electrons in the electromagnetic
field obtained from PIC simulations. The test electrons are assumed
to be injected near the cavity axis which passes through the cavity
centers. First we estimate the gain in the momentum of the electron
passing a one-period structure as a function of the initial field
phase $\Theta$ and $x$-coordinate of the electron 

\noindent 
\begin{eqnarray}
\Delta p_{y} & \simeq & \int_{0}^{T}eE_{y}dt,\label{wkb1}\\
\Delta p_{x} & \simeq & \int_{0}^{T}e\left(E_{x}-B_{z}\right)dt,\label{wkb2}
\end{eqnarray}
where $T=2\pi/\omega_{L}$ is the laser period. The gain in the momentum
of the electron travelling in the fields retrieved from PIC simulation
as function of $\Theta$ and $x$ is shown in Fig.\ref{fig3}. Electron
acceleration is efficient when the gain in the longitudinal momentum
peaks and the transverse momentum exhibits stable betatron oscillations.
It follows from Fig.\ref{fig3}a that the optimal regime of acceleration
corresponds to $\Theta\approx0$. In the ideal accelerator, the bunch
axis must coincide with the middle of the microcavity where the transverse
fields are zero. Numerical simulations show that the transverse force
is not exactly absent at the cavity symmetry axis due to higher mode
excitation and nonlinear interactions with plasma. As a result the
electron undergoes some betatron oscillations even if it starts to
move along the cavity symmetry axis. 

\begin{figure}
\includegraphics[width=0.5\columnwidth]{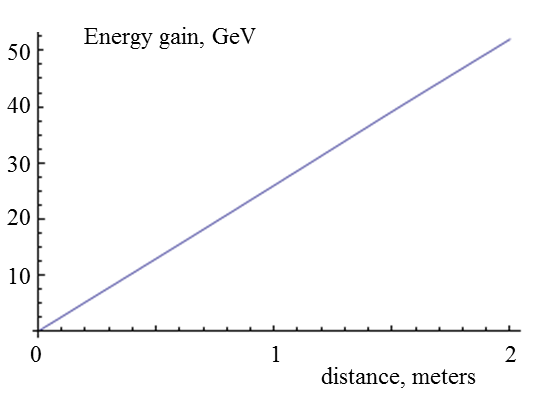}

\caption{The dependence of electron energy on the passing distance. The starting
parameters of the electron are $\gamma=200$, $x=0$ and $\Theta=-0.1\pi$.}

\label{fig4} 
\end{figure}

Using periodic boundary condition for the field distribution shown
in figure \ref{fig2-1} we analyze electron acceleration in a many-period
accelerating structure. The typical electron trajectories are shown
in figure \ref{fig3}b. It follows from the analysis that the stable
and efficient regime of electron acceleration corresponds to $\Theta\approx0$
that agrees with the estimates presented above. Long-term acceleration
is demonstrated in Fig.\ref{fig4}. The electron gains energy of about
$50\,$GeV over $2$ meters of the structure. The transverse betatron
amplitude of the accelerated electron remains less than $0.02\lambda$
during acceleration. 

\section{ELECTRON ACCELERATION IN OPEN PLASMA STRUCTURES}

The resonator microcavities have the advantage that they can trap
the laser pulse energy and amplify the field. However, it might be
difficult to achieve high finesses, and the resonating properties
are very sensitive to the plasma parameters, sizes of the microboxes
and the wall quality. In addition, the very thin semi-transparent
foil covering the microcavity can only survive very short pulses of
moderate intensity. Thus, non-resonating open structures have their
advantages.

\begin{figure}
\includegraphics[width=0.95\columnwidth]{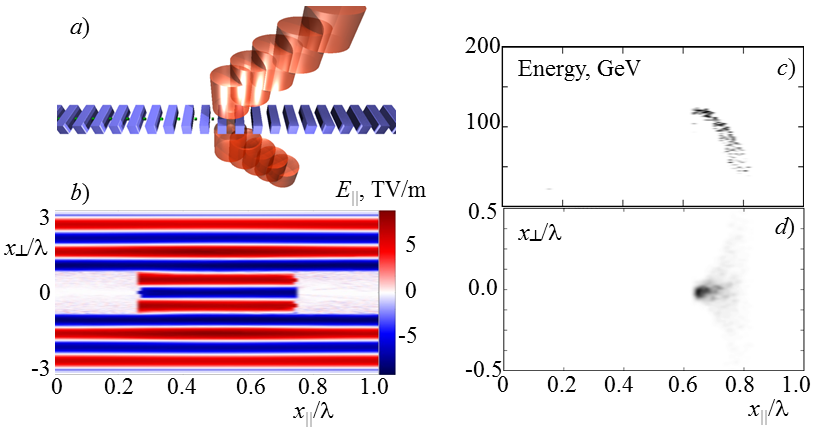}

\caption{a) Three-dimensional view of the accelerating structure: two counterpropagating
laser echelons hit a periodic plasma grid. b) The accelerating field
component. c) Longitudinal electron phase space after the accelerating
distance of 5cm. Electrons gained up to 120 GeV. d) Configuration
space of the electron bunch. The coma of the bunch where the highest
energy electrons are located has r.m.s. radius of about 20nm. }

\label{fig5} 
\end{figure}

Again, to satisfy a phase synchronism between a normally incident
laser pulse and an electron propagating in the transverse direction,
the plasma structure must have a spatial period defined by the relation
\eqref{eq:period}. Thus, for a relativistic electron, we need a structure
with the period equal to the laser wavelength $\lambda$. One may
think of different configurations. The simplest one is shown in Fig.\ref{fig5}a.
In this simulation we assume two counterpropagating phased laser echelons
interacting with a periodic plasma grid. The overdense plasma regions
have the length of $\lambda/2$ and are separated by vacuum regions
of the same length $\lambda/2$. The accelerating field configuration
is shown in Fig.\ref{fig5}b. We assumed that the lasers of the both
echelons have the peak relativistic amplitude of $eE_{L}/mc\omega=1$
so that the combined accelerating field in the center of the structure
reaches the peak amplitude $E_{{\rm ||}}=2E_{L}=7.5\,{\rm TV/m}$. 

In our simulations, we injected a continuous beam of test electrons
with initial $\gamma=200$ into the structure. About a quarter of
the electrons has been trapped and accelerated up to $120\,{\rm GeV}$
energy after a propagation distance of only $5.3\,{\rm cm}$, see
Fig.\ref{fig5}c. This corresponds to a sustained acceleration rate
of $2.3\,{\rm TeV/m}$. The configuration space of the electron bunch
is shown in Fig.\ref{fig5}d. Electrons with the highest energy are
also tightly focused and the coma of the bunch has an r.m.s. radius
of about $\sigma_{{\rm r.m.s.}}\approx20\,{\rm nm}$. 

The advantage of the open structure with laser echelons coming from
both sides is its robustness with respect to an exact position of
the plasma grid. The fields are defined by the laser relative phases
only. This makes the scheme unsensitive to e.g., mechanical jitter
of the grid.

We mention also another importat feature. The plasma plasma structure
remains invisible by the counterporpagating laser pulses. No modes
modes propagating along the plasma grid are generated and we see essentially
no perturbations in the laser field distribution in vacuum. This promises
a high quality acceleration with potentially high energy conversion
efficiency.

\begin{figure}
\includegraphics[width=0.6\columnwidth]{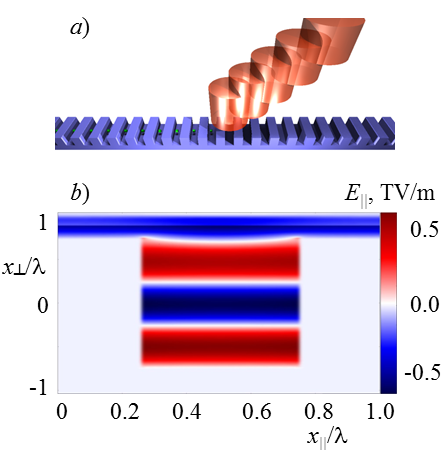}

\caption{a) Three-dimensional view of the accelerating structure: a single
laser echelon hits a periodic plasma grid on top of a plane substrate.
b) The accelerating field component. }

\label{fig6} 
\end{figure}

Yet, a very similar field distribution can be created with a single
echelon of laser pulses if we use a grid on a surface of a solid substrate,
see Fig.\ref{fig6}a. The accelerating field inside the cavities,
Fig.\ref{fig6}b, has a very similar structure as in Fig.\ref{fig5}b.
In this particular simulation we used a lower laser pulse amplitude
of of $eE_{L}/mc\omega=0.1$ so that the combined accelerating field
of the incident and reflected wave in the center of the cavity reaches
the peak amplitude $E_{{\rm ||}}\approx2E_{L}=0.7\,{\rm TV/m}$. Because
of the lower amplitude, the plasma response is more linear and less
noise due to excitation of nonlinear harmonics and higher modes is
seen in this picture. However, the exact position of the field maximum
in this scheme is defined no more by the laser phase, but rather by
the position of the reflecting surface of the substrate and thus is
fully sensitive to mechanical oscillations of the structure.

\section{DISCUSSION AND CONCLUSION}

We have considered a possibility to use an echelon of laser pulses
generated by the emerging CAN technology for direct particle acceleration
in periodic plasma structures. The advantage of the method is the
potentially extremely high acceleration rate above ${\rm TeV/m}$.
The reason is that the laser field itself is used and not a perturbative
plasma field (e.g. wake field). In our simulations, we show that a
sustained acceleration rate of $2.3\,{\rm TeV/m}$ is feasible with
the plasma structures. We discussed resonant and free streaming configurations.
The resonant plasma structures can trap energy of longer laser pulses
that would be harvested by the witness bunch. However, the resonant
structure is limited to moderate laser intensities of about $10^{14}\,{\rm W/cm^{2}}$
as the thin semi-transparent plasma foil is destroyed very fast. The
free streaming configurations can survive laser intensities above
$10^{18}\,{\rm W/cm^{2}}$ for several tens of femtoseconds. The plasma
structures survive a single shot only. However, due to their simplicity
and projected very low production costs, the structures can be replaced
for every laser shot at a kiloherz repetition rate. 
\begin{acknowledgments}
We acknowledge very enlightening discussions with Prof. G. Mourou
and Prof. T. Tajima. 

This work has been supported by the Government of the Russian Federation
(Project No. 14.B25.31.0008), by the Ministry of Science and Education
of the Russian Federation, the Russian Federal Program ``Scientific
and scientific-pedagogical personnel of innovative Russia'' (Agreement
N8835) \end{acknowledgments}


\begin{thebibliography}{References}
\bibitem{Mourou1985}D.~Strickland \& G.~Mourou, Opt. Commun. \textbf{56},
219 (1985).

\bibitem{Mangles2004} S.~P.~D.~Mangles, C.~D.~Murphy, Z.~Najmudin,
A.~G.~R.~Thomas, J.~L.~Collier, A.~E.~Dangor, E.~J.~Divall,
P.~S.~Foster, J.~G.~Gallacher, C.~J.~Hooker, Nature (London)
\textbf{431}, 535 (2004).

\bibitem{Geddes2004} C.G.R.~Geddes, C.~Toth, J.~van~Tilborg,
E.~Esarey, C.B.~Schroeder, D.~Bruhwiler, C.~Nieter, J.~Cary,
and W.P.~Leemans, Nature (London) \textbf{431}, 538 (2004).

\bibitem{Faure2004} J.~Faure, Y.~Glinec, A.~Pukhov, S.~Kiselev,
S.~Gordienko, E.~Lefebvre, J.-P.~Rousseau, F.~Burgy, and V.~Malka,
Nature (London) \textbf{431}, 541 (2004).

\bibitem{Leemans2006} W.~P.~Leemans, B.~Nagler, A.~J.~Gonsalves,
C.~Tih, K.~Nakamura, C.~G.~R.~Geddes, E.~Esarey, C.~B.~Schroeder,
and S.~M.~Hooker, Nat.~Phys. \textbf{2}, 696 (2006).

\bibitem{Downer2013} X.~Wang, R.~Zgadzaj, N.~Fazel, Z.~Li, S.
A. Yi, X.~Zhang,W.~Henderson, Y.-Y.~Chang, R..~Korzekwa, H.-E.~Tsai,
C.-H.~Pai, H.~Quevedo, G.~Dyer , E.~Gaul, M.~Martinez, A.~C.~Bernstein,
T.~Borger.~M.~Spinks, M.~Donovan, V.~Khudik, G.~Shvets, T.~Ditmire
\& M.~C.~Downer, Nature Comm. \textbf{4}, 1988 (2013).

\bibitem{Bubble}A. Pukhov, J. Meyer-ter-Vehn, Applied Physics B \textbf{74}
(4-5), 355-361 (2002).

\bibitem{Jeong2004}Y.~Jeong, J.~K.~Sahu, D.~N.~Payne \& J.~Nilsson,
Electron. Lett.\textbf{ 40}, 470 (2004). 

\bibitem{Liem2004}A.~Liem \textit{et al}., Conference on Lasers
and Electro-Optics \textbf{2}, 1067 (2004).

\bibitem{Ediam2011} T.~Eidam \textit{et al}., Opt. Express \textbf{19},
255 (2011).

\bibitem{Prawiharjo2008}J.~Prawiharjo \textit{et al}., Opt. Express
\textbf{16}, 15074 (2008).

\bibitem{Mourou2013}G.~Mourou, B.~Brocklesby, T.~Tajima and J.~Limpert,
Nature Photonics. \textbf{7}, 258 (2013).

\bibitem{Bellanger2010}C.~Bellanger \textit{et al}., Opt. Lett.
\textbf{35}, 3931 (2010).

\bibitem{Rosenzweig1995}J.~Rosenzweig, A.~Murokh, and C.~Pellegrini,
Phys.~Rev.~Lett. \textbf{74}, 2467 (1995).

\bibitem{Eddie2001}X.~Eddie Lin, Phys. Rev. ST Accel. Beams \textbf{4},
051301 (2001).

\bibitem{Huang1996}Y.~C.~Huang, D.~Zheng, W.~M.~Tulloch, and
R.~L.~Byer, Appl.~Phys.~Lett. \textbf{68}, 753 (1996).

\bibitem{Yoder2005}R.~B.~Yoder and J.~B.~Rosenzweig, Phys. Rev.
ST Accel. Beams \textbf{8}, 111301 (2005).

\bibitem{Tien1999}A.-C.~Tien, et al., Phys.~Rev.~Lett. \textbf{82},
3883 (1999).

\bibitem{Plettner2006}T.~Plettner, P.~P.~Lu, and R.~L.~Byer,
Phys.~Rev.~ST~Accel.~Beams \textbf{9} 111301 (2006).

\bibitem{VLPL}A. Pukhov Journal of plasma physics \textbf{61}, 425-433
(1999)

\bibitem{Shen2002}B.~Shen and M.~Y.~Yu, Phys.~Rev.~Lett. \textbf{89}
275004, (2002).\end{thebibliography}
\end{document}